\documentclass[12pt]{iopart}
\usepackage{epsfig}
\begin{document}

\title[Return or stock price differences]{Return or stock price differences}

\author{Jaume Masoliver~\footnote[1]{Corresponding author. E-mail: {\tt jaume@ffn.ub.es}}, Miquel Montero and Josep Perell\'o}

\address{Departament de F\'{\i}sica Fonamental, Universitat de Barcelona, Diagonal, 647, 08028-Barcelona, Spain}

\begin{abstract}

The analysis which assumes that tick by tick data is linear may lead to wrong conclusions if the underlying process is multiplicative. We compare data analysis done with the return and stock differences and we study the limits within the two approaches are equivalent. Some illustrative examples concerning these two approaches are given. Actual data is taken from S\&P 500 stock cash index.

\end{abstract}




\section{Introduction}

One of the most important problems in mathematical finance is to know the probability distribution of speculative prices. The first approach to the problem was given by Bachelier in 1900 when he modelled price dynamics as an ordinary random walk where prices can go up and down due to a variety of many independent random causes. Therefore, the distribution of prices has to be Gaussian~\cite{bachelier} due to the Central Limit Theorem: the sum of independent, or weakly dependent, random disturbances, all of them with finite variance, results in a Gaussian random variable.
 
Despite Bachelier's very early interest in stochastic modelling of stock prices, research on this topic is not again noticeable until 1930's. A renewed regard on financial markets appeared in the embryo school of American economists highly skilled in mathematics and statistics. In an ideal and theoretical framework, they believed that market was perfect in the sense that one cannot forecast future price changes based on past history alone. Therefore, they conclude that price changes have to be uncorrelated, and follow a Gaussian random process thus obeying the Central Limit Theorem enunciated above.

At that time, the main research in finance was addressed to test those theoretical hypothesis on real markets. In 1953, Kendall analyzed several American markets observing there that price changes behave like {\it wandering series} and discovering correlations in the price movements time series~\cite{kendall}. All of this partially contradicted the economic theory since, although Kendall confirmed the random nature of stock evolution, he also found correlations which were unacceptable in an ideal and perfect market framework.  

Few years later, Osborne tried to fit data with a Brownian motion model and looked for 
the form of the price changes empirical distribution~\cite{osborne}. Price movements modelled as a random walk implied that price can be negative with non-zero probability. In order to avoid the complications posed by the fact that stock prices must have a lower bound, he proposed to take the logarithm changes of prices instead of the price changes, {\it i.e.}, $\ln[S(t+\Delta)/S(t)]$ instead of $S(t+\Delta)-S(t)$. In this case, there were no need of limiting the process into a positive region. Osborne confronted new model with real markets and observed that the new variable (call it stock return, $R(t)=\ln[S(t)/S_0]$) evolved as a Gaussian random walk and had its increments uncorrelated. This feature was in concordance with the {\it perfect} market hypothesis, and gives more importance to the way data is taken from markets.

This historical introduction do not want to review the existent market models nor present a new model. From sixties to nowadays, we have much larger time series recorded from markets, and computers have allowed us to register the whole stock movements, the so-called tick by tick data or high-frequency data. People concern in financial markets have dramatically increased and there is a strong demand of high precision in the description of the speculative prices dynamics~\cite{lamoreux}. Data analysis in financial markets has thus become a relevant issue, and taking good estimators for checking economic theory and market models is an essential but delicate task. Kendall's 
and Osborne's works exemplify how important is to think over which data shall we handle and how we manipulate this data. Indeed, the way data is manipulated may lead to diverse and, in critical cases, contradictory conclusions.

More precisely, the purpose of this paper is to consider the way we analyze the financial market data. We focus in the differences between taking stock price and return increments, and we show their accuracy and range of validity for estimating parameters describing the market. Our intention is to display risk of obtaining wrong conclusions when we operate in an inadequate framework with the historical time series.
For instance, we see the different aspect adopted by the probability distribution in Figures~\ref{dif} and~\ref{ret} where we respectively take stock differences or return differences time as a data source. Database to make the comparison is the Standard \& Poor's 500 stock cash index ranging from 1983 to 1999, and we fulfill the assertions with a simple market model, the multiplicative Gaussian model which posseses properties also assumed in more sophisticated and realistic market models.
  
The paper is divided into six sections. In section~2 and~3 we present several functions related to the return and stock differences, respectively. We study their properties in general but also for the Wiener process particular case. In section~4 we show and compare the empirical probability distributions for S\&P 500 cash index for the stock and return differences. Section~5 concentrates on the estimators of the first and second moments. Conclusions are drawn in section~6.

\section{The return}

Characteristic functions are very useful for evaluating moments of any order. We depart from the stock return defined as the logarithm of the stock price, {\it i.e.}, $R(t)=\ln[S(t)/S_0]$, and we will thus calculate expressions related to the return stochastic variable. We will do it in order to obtain these expressions in terms of the return characteristic function~\cite{gardiner}.

The characteristic function is derived from the return probability density function (pdf) and defined as follows
\begin{equation}
\phi_R(\omega, t|0) = \int_{-\infty}^{\infty} dr \ \rme^{\rmi \omega r} 
\ p_R(r,t|0),
\label{phir}
\end{equation}
where the conditional density is defined as $p_R(r,t|0) \equiv p_R(r,t|r=0, t=0)$.

Return differences and the particular case assuming return evolution to be driven by a Wiener process are also studied. All these calculations are done with the demand that process is Markovian and homogeneous.
  
\subsection{Some expressions for the return stochastic variable}

Let us present some results related to the return variable using equation~(\ref{phir}). For instance, the first moment of the return is
\begin{equation}
\langle R(t) \rangle = -\rmi \ \partial_{\omega} \phi_R(\omega,t|0)
\Bigr|_{\omega=0},
\label{<r>}
\end{equation}
and the second moment is
\begin{equation}
\langle R(t)^2 \rangle = - \ \partial_{\omega \omega}^2 
\phi_R(\omega,t|0) \Bigr|_{\omega=0}.
\label{<r^2>}  
\end{equation}
We can also derive the variance
\begin{equation}
{\rm Var} [R(t)] = \langle R(t)^2 \rangle -\langle R(t)\rangle^2 =
 -\ \partial_{\omega \omega}^2 
\phi_R(\omega,t|0) + \left[\partial_{\omega} \phi_R(\omega,t|0) \right]^2
\Bigr|_{\omega=0},
\label{varr}
\end{equation}
where equations~(\ref{<r>}) and~(\ref{<r^2>}) have been taking into account.

On the other hand, we can obtain the joint probability density function at two different times with the condition that underlying process for the return is Markovian, {\it i.e.},
\begin{equation}
p_R(r_1,t_1;r_2,t_2|0)= p_R(r_1,t_1|r_2,t_2) p_R(r_2,t_2|0),
\label{markovian}
\end{equation}
whenever $t_1 \geq t_2$. But, if we also impose that process is homogeneous in time and return, {\it i.e.}, 
\begin{equation}
p_R(r_1,t_1|r_2,t_2)= p_R(r_1-r_2,t_1-t_2|0),
\label{homogeneous}
\end{equation}
We can go further and see that
\begin{eqnarray*} 
\fl \phi_R(w_1,t_1;w_2,t_2|0) &=&
\int_{-\infty}^{\infty} dr_1 \ \rme^{\rmi \omega_1 r_1} 
\int_{-\infty}^{\infty} dr_2 \ \rme^{\rmi \omega_2 r_2} \
p_R(r_1,t_1;r_2,t_2|0) 
\\ &=&
\int_{-\infty}^{\infty} dr_1 \int_{-\infty}^{\infty}
 dr_2 \ \rme^{\rmi (\omega_1 r_1+\omega_2 r_2)}
\ p_R(r_1-r_2,t_1-t_2|0) \ p_R(r_2,t_2|0), 
\end{eqnarray*}
and we thus obtain the joint characteristic expressed as a product of two characteristic functions. That is:
\begin{equation}
\phi_R(w_1,t_1;w_2,t_2|0)=\phi_R(\omega_1,t_1-t_2|0) \ \phi_R(\omega_1 + \omega_2,t_2|0).
\label{phirr}
\end{equation}

Indeed, the correlation function for the return can be written in terms of the joint characteristic function. Using equation~(\ref{phirr}), we can derive the correlation function which reads
\begin{eqnarray*}
\fl
\langle R(t_1)R(t_2) \rangle &=& -
\partial^2_{w_1 w_2} \phi_R(\omega_1,t_1;\omega_2,t_2|0)
\Bigr|_{\omega_1,\omega_2=0}
\\ &=&
- \left[ \partial^2_{\omega \omega} \phi_R(\omega,t_2|0)
+
\partial_\omega \phi_R(\omega,t_2|0)
\ \partial_\omega \phi_R(\omega,t_1-t_2|0) \right]\Bigr|_{\omega=0}.
\end{eqnarray*}
Taking into account equations~(\ref{<r>}) and~(\ref{<r^2>}), we write an expression for the
correlation function in terms of the first and second moments of $R$. Thus,
\begin{equation}
\langle R(t_1)R(t_2) \rangle =
\langle R(t_2)^2 \rangle + \langle R(t_2) \rangle \langle R(t_1-t_2) \rangle
\qquad (t_1 \geq t_2).
\label{corrr}
\end{equation}

It is usually defined a coefficient which evaluates the degree of correlation between a pair stochastic quantities~\cite{cramer}. The coefficient $\rho$ here defined is enclosed between the interval $\rho = [-1,1]$. In case that $\rho=0$, it is said that 
the pair of stochastic quantities are {\it uncorrelated}. And in any other case, we shall say that quantities are {\it correlated} and that the correlation is {\it positive} or {\it negative} according as $\rho >0$ or $\rho <0$. When the coefficient raises one of its extreme values, it is said that one quantity is a linear function of the other, and the two quantities vary in the same linear sense, $\rho=1$, or in inverse sense, $\rho=-1$. For the case referred to the stochastic return variable, 
the correlation coefficient reads
\begin{equation}
\rho (t_1,t_2)\equiv \frac{\langle R(t_1)  R(t_2) \rangle -
 \langle R(t_1) \rangle \langle R(t_2) \rangle}{\sqrt{
{\rm Var}[R(t_1)] \ {\rm Var}[R(t_2)]}}.
\label{rho}
\end{equation}
We can simplify this expression with the help of equations~(\ref{varr}) and~(\ref{corrr})
\[
\rho (t_1,t_2)= \frac{\langle R(t_2)^2 \rangle +
\langle R(t_2)\rangle \langle R(t_1-t_2)\rangle
-\langle R(t_1) \rangle \langle R(t_2) \rangle}
{\sqrt{{\rm Var}[R(t_1)] \ {\rm Var}[R(t_2)]}}.
\]
We need to know an equivalent expression for $ \langle R(t_1-t_2)\rangle$
\begin{eqnarray*}
\langle R(t_1-t_2) \rangle &=&
\int_{-\infty}^{\infty}da \ a \ p_R(a,t_1-t_2|0)
\\
&=&
\int_{-\infty}^{\infty}dr_2 \int_{-\infty}^{\infty}da \ a \ p_R(a,t_1-t_2|0)
\ p_R(r_2,t_2|0),
\end{eqnarray*}
where we have only  added an expression which value is one due to the fact $p(r_2,t_2|0)$ is normalized. Taking into account that process is Markovian and 
homogeneous whose definitions are given by equations~(\ref{markovian})--(\ref{homogeneous})
and doing the change of variables $a=r_1-r_2$, we have 
\begin{eqnarray*}
\langle R(t_1-t_2) \rangle
&=& 
\int_{-\infty}^{\infty}dr_2 \int_{-\infty}^{\infty}dr_1 \ (r_1-r_2) 
\ p_R(r_1,t_1|r_2,t_2)\ p_R(r_2,t_2|0)
\\
&=& 
\int_{-\infty}^{\infty}dr_2 \int_{-\infty}^{\infty}dr_1 \ (r_1-r_2) 
\ p_R(r_1,t_1;r_2,t_2|0).
\end{eqnarray*}
Hence,
\begin{equation}
\langle R(t_1-t_2) \rangle = \langle R(t_1)-R(t_2) \rangle=
\langle R(t_1) \rangle - \langle R(t_2) \rangle.
\label{<r1-r2>}
\end{equation}
After simple manipulations we finally obtain
\begin{equation}
\rho(t_1,t_2)
= \sqrt{\frac{{\rm Var}[R(t_2)]}{{\rm Var}[R(t_1)]}}.
\label{rhor}
\end{equation}

\subsection{The stock return difference}

However, the variable in which we are specially interested is the one referred to the return differences. We define a new stochastic variable called stock return differences by 
\begin{equation}
W(t;\tau)\equiv R(t+\tau)-R(t).
\label{w}
\end{equation}

This stochastic variable has the same pdf and, therefore, same moments and correlation function as the return. Let us show this. From equation~(\ref{w}), we see that
\begin{eqnarray*}
p_W(w,t;\tau) &=& 
\int_{-\infty}^{\infty} dr \int_{-\infty}^{\infty} dr' \
\delta[w-(r-r')] \ p_R(r,t+\tau;r',t|0)
\\
&=& 
\int_{-\infty}^{\infty} dr \int_{-\infty}^{\infty} dr' \
\delta[w-(r-r')] \ p_R(r-r',\tau|0) \ p_R(r',t|0),
\end{eqnarray*}   
where we have taken into account equations~(\ref{markovian}) and~(\ref{homogeneous}). Implementing the delta function in the inner integral of this expression we get 
\[
p_W(w,t;\tau)=p_R(w,\tau|0) \int_{-\infty}^{\infty}dr \ p_R(r-w,t|0),
\]
and since $p_R(r,t|0)$ is normalized we finally obtain
\begin{equation}
p_W(w,t;\tau)=p_R(w,\tau|0),
\label{pw}
\end{equation}
which shows that $p_W$ is identical to the return pdf $p_R$. Observe that distribution is only function of the time difference $\tau$ and does not depend on time $t$. Therefore, we can take expressions presented in equations~(\ref{<r>})--(\ref{rhor}) and replace $t$ by $\tau$ in order to give the equivalent expressions for the return differences $W(t;\tau)$.

We now study the autocorrelation between the variable $W$ evaluated at distinct times $t$ and $t'\geq t+\tau$. Thus,
\begin{eqnarray*}
\langle W(t;\tau) W(t';\tau)\rangle &=& \langle R(t+\tau) R(t'+\tau)\rangle +\langle R(t) R(t')\rangle \\ &&
- \langle R(t+\tau) R(t')\rangle -\langle R(t) R(t'+\tau)\rangle,
\end{eqnarray*}
where we have decomposed the function $W$ in terms of the return $R$. Equation~(\ref{corrr}) gives us the value of each autocorrelation. Taking into account the requeriment that $t'\geq t+\tau$ and after simple manipulations, we get
\begin{eqnarray*}
\langle W(t;\tau) W(t';\tau)\rangle &=& \langle R(t+\tau)\rangle \langle R(t'-t)\rangle +\langle R(t) \rangle \langle R(t'-t)\rangle \\ && - \langle R(t+\tau) \rangle \langle R(t'-t-\tau)\rangle -\langle R(t)\rangle \langle R(t'+\tau-t)\rangle.
\end{eqnarray*}
We can go one step further with the help of equation~(\ref{<r1-r2>}) which gives the the time invariant property for the first moment of the return. In terms of the return differences, the autocorrelation reads 
\[
\langle W(t;\tau) W(t';\tau)\rangle=\langle W(t;\tau)\rangle \langle W(t';\tau)\rangle,
\]
and thus see that the $W(t;\tau)$ quantities are uncorrelated.

In addition, we observe that the returns increments and the return itself are uncorrelated stochastic variables since the correlation function is 
\[
\langle W(t;\tau) R(t) \rangle =
\langle R(t+\tau) R(t) \rangle-\langle R(t)^2 \rangle = \langle R(\tau) \rangle \langle R(t) \rangle,
\]
where equation~(\ref{corrr}) is been applied. And we may write
\begin{equation}
\langle W(t;\tau) R(t) \rangle = 
\langle W(t;\tau) \rangle \langle R(t) \rangle.
\label{corrwr}
\end{equation}
thus being zero their correlation coefficient defined in equation(\ref{rho}). We therefore conclude that $W(t;\tau)$ and $R(t)$ are uncorrelated. This is not surprising since Markovian condition and homogeneity leads directly to the statement that $W(t;\tau)$ and $R(t)$ are independent stochastic variables. Indeed, if we take the joint pdf, we can first implement Markovian condition~(\ref{markovian}) 
\[
p_R(r',t';r,t)=p_R(r'-r,t'-t) p_R(r,t),
\]
and, afterwards, homegeneity~(\ref{homogeneous}) let us write joint pdf in the form
\[
p_R(r',t';r,t)=p_W(w,\tau) p_R(r,t).
\]
From last equation, we there see that joint pdf becomes the independent product of two other pdfs which are, in effect, the distributions of $W(t,\tau)$ and $R(t)$.

\subsection{A simple model for the return: The Wiener process} 

Let us now study these expressions for the Wiener process with drift $\mu$
and diffusion coefficient $\sigma$, a well-known market 
model in the literature~\cite{osborne}. The conditional pdf reads
\begin{equation}
p_R(r,t|0)= \frac{1}{\sqrt{2 \pi \sigma^2 t}} 
\exp\left[-\frac{(r-\mu t)^2}{2 \sigma^2 t}\right].
\label{wp}
\end{equation}
and the characteristic function defined in equation~(\ref{phir}) is
\begin{equation}
\phi_R(w,t|0)=
\exp\left[\left(\rmi \mu \omega -\frac{1}{2}\sigma^2 \omega^2\right) t\right].
\label{wphir}
\end{equation}
From this we see that
\begin{equation}
\langle R(t) \rangle= \mu t, \qquad 
\langle R(t)^2 \rangle= \sigma^2 t + \mu^2 t^2,
\qquad {\rm Var}[R(t)] = \sigma^2 t.
\label{w<r>}
\end{equation}
Moreover, we can also calculate the joint characteristic function
obtained with the help of equations~(\ref{phirr}) and~(\ref{wphir}). Thus,
\begin{eqnarray}
\fl
\phi_R(w_1,t_1;w_2,t_2|0) &=&
\exp\left\{\left(\rmi \mu \omega_1-
\frac{1}{2}\sigma^2 \omega_1^2\right)(t_1-t_2)\right\}
\nonumber \\ &&
\times
\exp\left\{\left[\rmi \mu (\omega_1+\omega_2) -
\frac{1}{2}\sigma^2 (\omega_1+\omega_2)^2\right]t_2\right\}
\nonumber \\
&=& \exp\left[\rmi \mu (\omega_1 t_1+\omega_2 t_2) -\frac{1}{2}\sigma^2 (\omega_1^2 t_1 +\omega_2^2 t_2+ 2 \omega_1 \omega_2 t_2)\right]. 
\nonumber \\
\label{wphirr}
\end{eqnarray}
Then, the correlation function according to equation~(\ref{corrr}) is
\begin{equation}
\langle R(t_1)R(t_2) \rangle = \sigma^2 t_2 +\mu^2t_2 t_1,
\label{wcorr}
\end{equation}
and the correlation coefficient can be obtained with 
equation~(\ref{rhor}) once we know the variance of equation~(\ref{w<r>}).
Hence,
\begin{equation}
\rho(t_1,t_2)= \frac{t_2}{t_1},
\label{wrhor}
\end{equation}
where since $t_1\geq t_2$ the coefficient $\rho$ is positive and less than 1.

As shown in general, the expressions for the return differences
in the Wiener case are equivalent to those presented for the return.
We note that, for this case, return differences and return stochastic 
variables are also uncorrelated (see equation~(\ref{corrwr})).

\section{The stock share price}

The purpose of this section is to derive the functions related to the 
stock price stochastic variable $S(t)$ 
in terms of the results obtained for the return $R(t)$.
The section is analogous to the one of the return variable but 
we now implement the same equations to the stock price stochastic variable.
We will also study price differences stochastic variable and the particular
Wiener process as a market model.

\subsection{Some expressions for the stock price stochastic variable}

We remember the relation between the return and the stock price:
\[ 
S=S_0 \ \rme^R.
\]
Our intention is to derive several expressions related to the stock 
price and write them, at the end, in terms of the characteristic function of 
the return.
Therefore, the first moment and the second moment are
respectively
\begin{equation}
\langle S(t) \rangle = S_0 \ \langle \rme^{R(t)} \rangle = S_0 \ \phi_R(-\rmi,t|0), 
\label{<s>}
\end{equation} 
and
\begin{equation}
\langle S(t)^2 \rangle = S_0^2 \ \langle \rme^{2R(t)} \rangle =
S_0^2 \ \phi_R(-2\rmi,t|0).
\label{<s^2>}
\end{equation}
The variance thus reads
\begin{equation}
{\rm Var}[S(t)]=  S_0^2 \left[ 
\langle \rme^{2R(t)} \rangle - \langle \rme^{R(t)} \rangle^2 \right]
= S_0^2 \left[ \phi_R(-2\rmi,t|0) - \phi_R(-\rmi,t|0)^2 \right].
\end{equation}

In addition, the characteristic function for the stock price can also
be obtained as a sum of characteristic functions for the return. That is:
\begin{eqnarray}
\fl \phi_S(w,t|S_0) &=& \int_0^{\infty} ds \ \rme^{\rmi \omega s} p_S(s,t|S_0)
= \int_{-\infty}^{\infty} dr \ \rme^{\rmi \omega S_0 \rme^r} p_R(r,t|0)
\nonumber \\
&=&
\sum_{n=0}^{\infty} \frac{(\rmi \omega S_0)^n}{n!}  
\int_{-\infty}^{\infty} dr \ p_R(r,t|0) \ \rme^{nr} =
\sum_{n=0}^{\infty} \frac{(\rmi \omega S_0)^n}{n!} \ \phi_R(-\rmi n, t|0).
\label{phis}
\end{eqnarray}
We also see from equation(\ref{phirr}) that, when $t_1\geq t_2$,
\begin{eqnarray}
\langle S(t_1)S(t_2) \rangle &=& \int_{-\infty}^{\infty}dr_1
\int_{-\infty}^{\infty}dr_2 \ S_0^2 \ \rme^{r_1+r_2}
\ p_R(r_1,t_1;r_2,t_2|0) \nonumber \\
&=& 
S_0^2 \ \phi_R(-\rmi,t_1-t_2|0) \ \phi_R(-2\rmi,t_2|0),
\label{corrs}
\end{eqnarray}
from where we can write autocorrelation in terms of first and second moments if we take into account equations~(\ref{<s>}) and~(\ref{<s^2>}). That is:
\begin{equation}
\langle S(t_1)S(t_2) \rangle= \langle S(t_1-t_2) \rangle \frac{\langle S(t_2)^2 \rangle}{S_0},
\label{corrs1}
\end{equation}
and the correlation coefficient defined in equation(\ref{rho}) is
\begin{equation}
\fl
\rho(t_1,t_2)=\frac{\phi_R(-\rmi,t_1-t_2|0) \ \phi_R(-2\rmi,t_2|0)-
\phi_R(-\rmi,t_1|0) \ \phi_R(-\rmi,t_2|0)}
{\sqrt{\left[\phi_R(-2\rmi,t_1|0) - \phi_R(-\rmi,t_1|0)^2 \right] \ \left[
\phi_R(-2\rmi,t_2|0) - \phi_R(-\rmi,t_2|0)^2\right]}}.
\label{rhos}
\end{equation}

\subsection{The stock price differences}
The stock price differences are actually an usual variable involved in data analysis. We can define the stock price difference by
\begin{equation}
Z(t;\tau) \equiv S(t+\tau)-S(t),
\label{Z}
\end{equation}
or the relative stock price difference in the following way
\begin{equation}
Y(t;\tau) \equiv \frac{S(t+\tau)-S(t)}{S(t)}.
\label{Y}
\end{equation}

In consequence, we can do a similar analysis to these variable as the one already done for the stock. We want to obtain the several expressions for these variables in terms of the return characteristic function. 

For instance, the pdf of $Z(t;\tau)$ is related to the stock price pdf as
\[
p_Z(z,t;\tau)= \int_{0}^{\infty} ds \int_{0}^{\infty} ds' \
\delta[z-(s-s')] \ p_S(s,t+\tau;s',t|S_0),
\]
and the characteristic function is
\begin{eqnarray}
\phi_Z(\omega,t;\tau)&=&\int_{0}^{\infty}ds \int_{0}^{\infty}ds'
\rme^{\rmi \omega (s-s')} p_S(s,t+\tau;s',t|S_0)  \nonumber \\
&=&
\int_{-\infty}^{\infty}dr\int_{-\infty}^{\infty}dr'
\rme^{\rmi \omega S_0(\exp r-\exp r')} p_R(r,t+\tau;r',t|0) \nonumber \\
&=&
\int_{-\infty}^{\infty}dr \int_{-\infty}^{\infty}dr'
\ \rme^{\rmi \omega S_0(\exp r-\exp r')} \ p_R(r-r',\tau|0) \ p_R(r',t|0),
\nonumber
\end{eqnarray}
where Markovian and homogeneity conditions have been implemented. Now, as before, we can expand the exponentials in order to give an expression in terms of the return characteristic function. Thus,
\begin{eqnarray}
\fl \phi_Z(\omega,t;\tau)
&=&
\sum_{n=0}^{\infty} \frac{(\rmi \omega S_0)^n}{n!} 
\sum_{k=0}^{n} \left( 
\begin{array}{c}
n \\ k 
\end{array}
\right)
(-1)^k 
\int_{-\infty}^{\infty}dr' \rme^{nr'} p_R(r',t|0) 
\nonumber \\
&& \; \; \; \; \; \; \; \; \;\; \; \; \;\; \; \; \;\; \;\; \; \; \; \; \; \; \;\; \; \; \; \; \; \;\; \;  \; \; \;\; \; \; \; \times \int_{-\infty}^{\infty}dr 
\rme^{k(r-r')} p_R(r-r',\tau|0) \nonumber \\
&=&
\sum_{n=0}^{\infty} \frac{(\rmi \omega S_0)^n}{n!} 
\sum_{k=0}^{n} \left(
\begin{array}{c}
n \\ k
\end{array}
\right)
(-1)^k  \phi_R(-\rmi n,t|0) \ \phi_R(-\rmi k,\tau|0).
\label{phiz}
\end{eqnarray}
 
Hence, its first moment is obtained with the definition in
equation~(\ref{<r>})
\begin{equation}
\langle Z(t;\tau) \rangle = S_0 \ \phi_R(-\rmi,t|0) \ [\phi_R(-\rmi,\tau|0)-1],
\label{<z>}
\end{equation}
but the second moment needs a longer calculation with the following final expression
\begin{equation}
\langle Z(t;\tau)^2 \rangle=
S_0^2 \ \phi_R(-2 \rmi,t|0) \ [1+\phi_R(-2 \rmi,\tau|0) -
2 \ \phi_R(-\rmi,\tau|0)].
\label{<z^2>}
\end{equation}
And we can also present the variance
\begin{eqnarray}
\fl {\rm Var}[Z(t;\tau)] = S_0^2 \ \left\{ 
\phi_R(-2\rmi,t|0) \ [1+\phi_R(-2 \rmi,\tau|0) -
2 \ \phi_R(-\rmi,\tau|0)] \right.
\nonumber \\
 \left.
\ \ \ \ \
-\ \phi_R(-\rmi,t|0)^2 \ [1+\phi_R(-\rmi,\tau|0)^2 -
2 \ \phi_R(-\rmi,\tau|0)] \right\}.
\label{varz}
\end{eqnarray}
Finally, we can study the correlation between stock differences and the stock itself. The correlation can be derived with equation~(\ref{corrs}) and results 
\[
\langle Z(t;\tau) S(t) \rangle =
\langle S(t+\tau)S(t) \rangle - \langle S(t)^2 \rangle
= S_0^2  \ \phi_R(-2\rmi,t|0) [\phi_R(-\rmi,\tau|0) - 1].
\]
Simple manipulations that take into account equation~(\ref{<z>}) let us write
\begin{equation}
\langle Z(t;\tau) S(t) \rangle = \frac{\langle S(t)^2 \rangle}{\langle S(t)\rangle}
\langle Z(t;\tau)\rangle.
\label{corrzs}
\end{equation}
And therefore the correlation coefficient is nonzero as it happens with the correlation coefficient between $W(t;\tau)$ and $R(t)$ presented in equation~(\ref{corrwr}).

On the other hand, we can perform an equivalent analysis but for the relative stock difference, $Y(t;\tau)$, given by equation~(\ref{Y}). Thus the characteristic function is
\begin{equation}
\phi_Y(\omega,t;\tau)
= \rme^{-\rmi\omega} \ \sum_{n=0}^{\infty} \frac{(\rmi\omega)^n}{n!} \phi_R(-\rmi n,\tau|0).
\label{phiY}
\end{equation}
Observe that characteristic function for $Y(t;\tau)$ is very similar to the one for the stock~(\ref{phis}) but now the function is evaluated at time $\tau$, $S_0$ does not appear, and an imaginary exponential appears in equation~(\ref{phiY}). This is not surprising since, due to the independency on time $t$ in equation~(\ref{phiY}), the variable $Y$ can be expressed as
\[
Y(\tau)=S(\tau)/S_0-1
\]
thus implying the properties above mentioned. First and second moments, and variance are
\begin{equation}
\langle Y(t;\tau) \rangle = \phi_R(-\rmi,\tau|0)- 1,
\label{<y>}
\end{equation}
\begin{equation}
\langle Y(t;\tau)^2 \rangle=  1 + \phi_R(-2\rmi,\tau|0) -
2 \ \phi_R(-\rmi,\tau|0),
\label{<y^2>}
\end{equation}
\begin{equation}
{\rm Var} [Y(t;\tau)] =  \phi_R(-2\rmi,\tau|0) - \phi_R(-\rmi,\tau|0)^2.
\label{vary}
\end{equation}
In addition, the correlation between $Y(t;\tau)$ and $S(t)$ is
\begin{equation}
\langle Y(t;\tau) S(t) \rangle = \langle Z(t;\tau) \rangle,
\label{corrys}
\end{equation}
but from equations~(\ref{<s>}),~(\ref{<z>}), and~(\ref{<y>}) we get
\begin{equation}
\langle Z(t;\tau)\rangle=\langle S(t) \rangle \langle Y(t;\tau) \rangle.
\label{<w>1}
\end{equation}
Therefore, $S(t)$ and $Y(t;\tau)$ are uncorrelated stochastic variables.
 
\subsection{The stock variables in the Wiener case}

We take into account the results of the two last subsections and characteristic
function of the return for the Wiener case of equation~(\ref{wphir}).

For the stock itself we can derive the following expressions. 
Therefore, we will subsequently obtain 
the first moment for the stock
\begin{equation}
\langle S(t) \rangle = S_0 \ \rme^{\left(\mu+\frac{1}{2}\sigma^2\right) t},
\label{w<s>}
\end{equation}
where it can be easily seen from equations~(\ref{w<s>}) and~(\ref{<s>}) that 
\[
\ln \left[ \frac{\langle S(t) \rangle}{S_0} \right]=
\left(\mu+\sigma^2/2 \right) t
\neq \mu t = \langle R(t) \rangle,
\]
due to what is called the spurious drift component. Indeed, we can obtain the second moment which is
\begin{equation}
\langle S(t)^2 \rangle = S_0^2 \ \rme^{2(\mu + \sigma^2) t}.
\label{w<s^2>}
\end{equation}

In addition, it is possible to express the characteristic function as follows
\begin{equation}
\phi_S(\omega, t|S_0)= 
\sum_{n=0}^{\infty} \frac{(i \omega S_0)^n}{n!} \
\rme^{(n \sigma)^2 t/2} \ \rme^{\mu n t},
\label{wphis}
\end{equation}
and the correlation function for the stock as
\begin{eqnarray}
\langle S(t_1)S(t_2) \rangle
&=&
S_0^2 \  \rme^{\left(\mu+\frac{1}{2}\sigma^2\right) (t_1-t_2)}
\ \rme^{2(\mu + \sigma^2) t_2}
\nonumber \\
&=&
S_0^2 \  \rme^{\mu(t_1+t_2)} \ \rme^{\frac{1}{2}\sigma^2(t_1+3t_2)},
\label{wcorrs}
\end{eqnarray}
where we take into account equations~(\ref{corrs}) 
and~(\ref{w<s>})--(\ref{w<s^2>}).
And, on the other hand, we have
the characteristic function for the price differences
\begin{equation}
\phi_Z(\omega,t;\tau)
=
\sum_{n=0}^{\infty} \frac{(\rmi \omega S_0)^n}{n!} 
\rme^{(n \sigma)^2 t/2} \ \rme^{\mu n t} \
\sum_{k=0}^{n} \left(
\begin{array}{c}
n \\ k
\end{array}
\right)
(-1)^k  
\rme^{(k \sigma)^2 t/2} \ \rme^{\mu k t}.
\label{wphiz}
\end{equation}
And we can derive the first moment
\begin{equation}
\langle Z(t;\tau) \rangle = S_0 \ 
\rme^{(\mu+\sigma^2/2)t} \ [\rme^{(\mu+\sigma^2/2)\tau}-1],
\label{w<z>}
\end{equation}
and second moment
\begin{equation}
\langle Z(t;\tau)^2 \rangle = S_0 \ 
\rme^{2(\mu+\sigma^2)t} \ [1+\rme^{2(\mu+\sigma^2)\tau}-
2 \rme^{(\mu+\sigma^2/2)\tau}].
\label{w<z^2>}
\end{equation}

We can also make similar calculus but for the relative difference.
The characteristic function is almost the same as the one for the stock:
\begin{equation}
\phi_Y(\omega, t|S_0)= \rme^{-\rmi\omega} 
\sum_{n=0}^{\infty} \frac{(\rmi \omega)^n}{n!} \ 
\rme^{(n \sigma)^2 t/2} \ \rme^{\mu n t},
\label{wphiy}
\end{equation}
and the first and second moments are respectively
\begin{equation}
\langle Y(t;\tau) \rangle = \rme^{\left(\mu+\sigma^2/2 \right)\tau}- 1,
\label{w<y>}
\end{equation}
\begin{equation}
\langle Y(t;\tau)^2 \rangle=  1 + \rme^{\left(\mu+\sigma^2 \right) 2 \tau} -
2 \ \rme^{\left(\mu+\sigma^2/2 \right) \tau}.
\label{w<y^2>}
\end{equation}
The first moment of $Y(t;\tau)$ is directly related to the first moment
of the stock $S(\tau)$ but differs from the $W(t;\tau)$. However, when $\tau$ is small, we thus have (up to first order)
\[
 \langle Y(t;\tau) \rangle \sim
\left(\mu+\sigma^2/2 \right) \tau
\neq \mu \tau = \langle W(t;\tau) \rangle,
\]
and, similarly, the second moment 
\[
\langle Y(t;\tau)^2 \rangle  
\sim \langle W(t;\tau)^2 \rangle \sim \sigma^2 \tau.
\]
The main difference between the two averages is the spurious drift component which appears in the first moment of $Y(t;\tau)$. We do not present a similar analysis on averages over $S(t)$, since calculus leads us to complex expressions depending on time $t$. This feature also present in the $Z(t;\tau)$ case and we will show in Section~5 that this fact is empirically observed in real markets.   

\section{Empirical probability distributions}

\begin{figure}
\begin{center}
\epsfbox{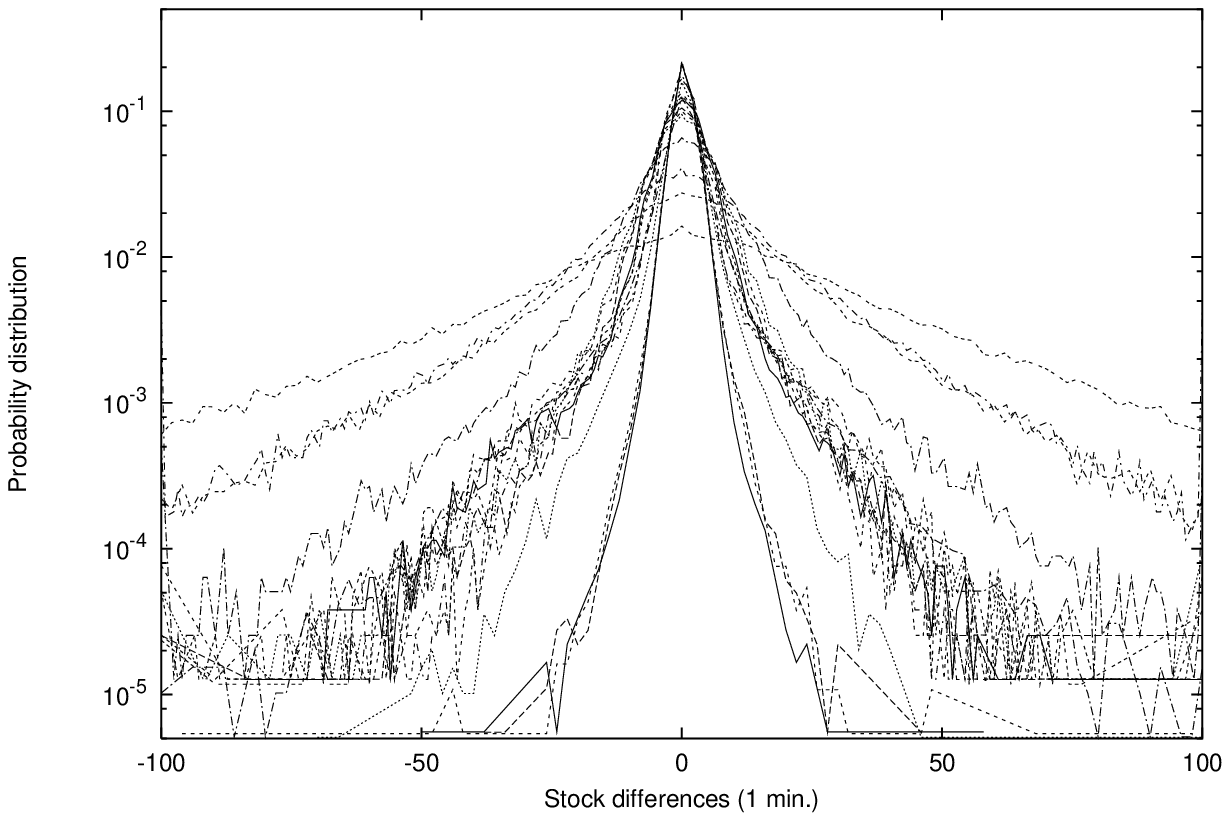}
\epsfbox{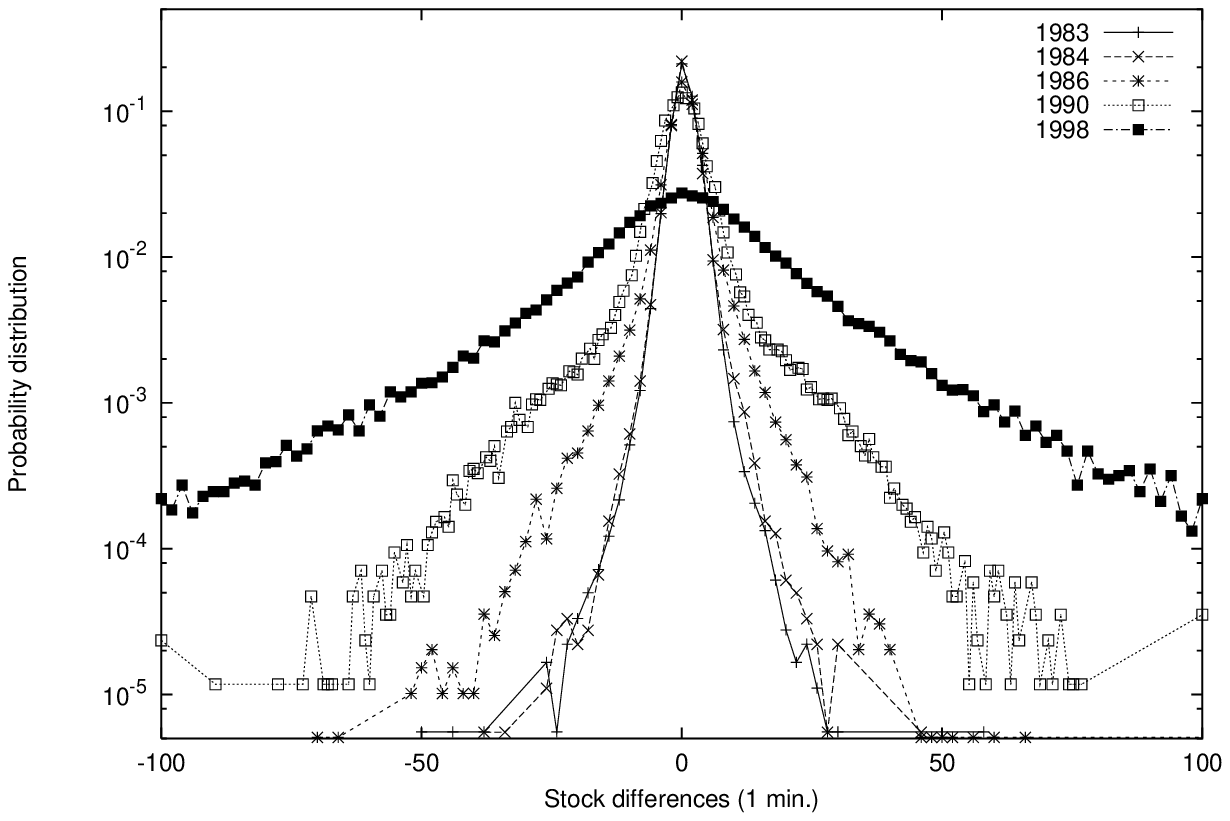}
\end{center}
\caption{We show the empirical pdf's for the tick data of the Standard \& Poor's 500 stock cash index differences. First graph involves one-minute stock differences for years ranging from 1983 to 1999. Second graph is a detail of the previous graph plotting pdf's of for years exponentially distributed between 1983 and 1998.}
\label{dif}
\end{figure}

\begin{figure}
\begin{center}
\epsfbox{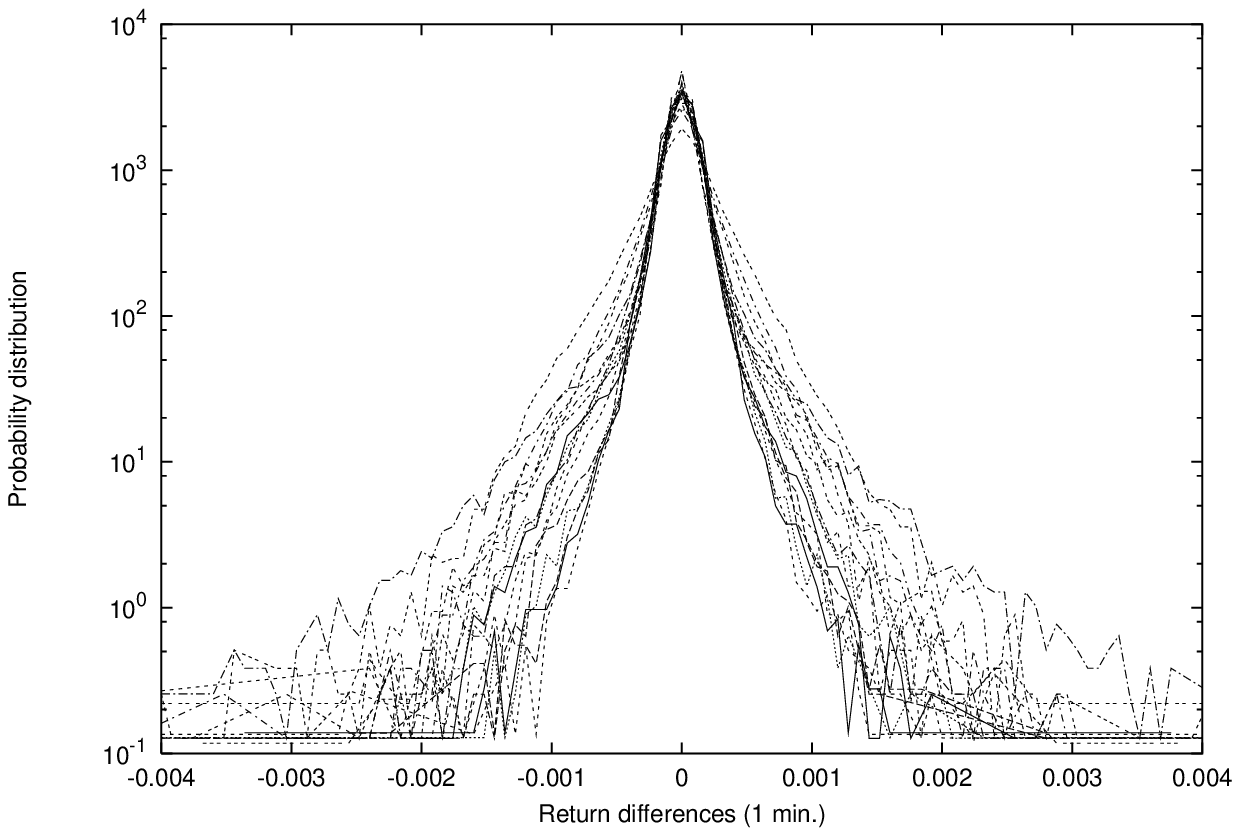}
\epsfbox{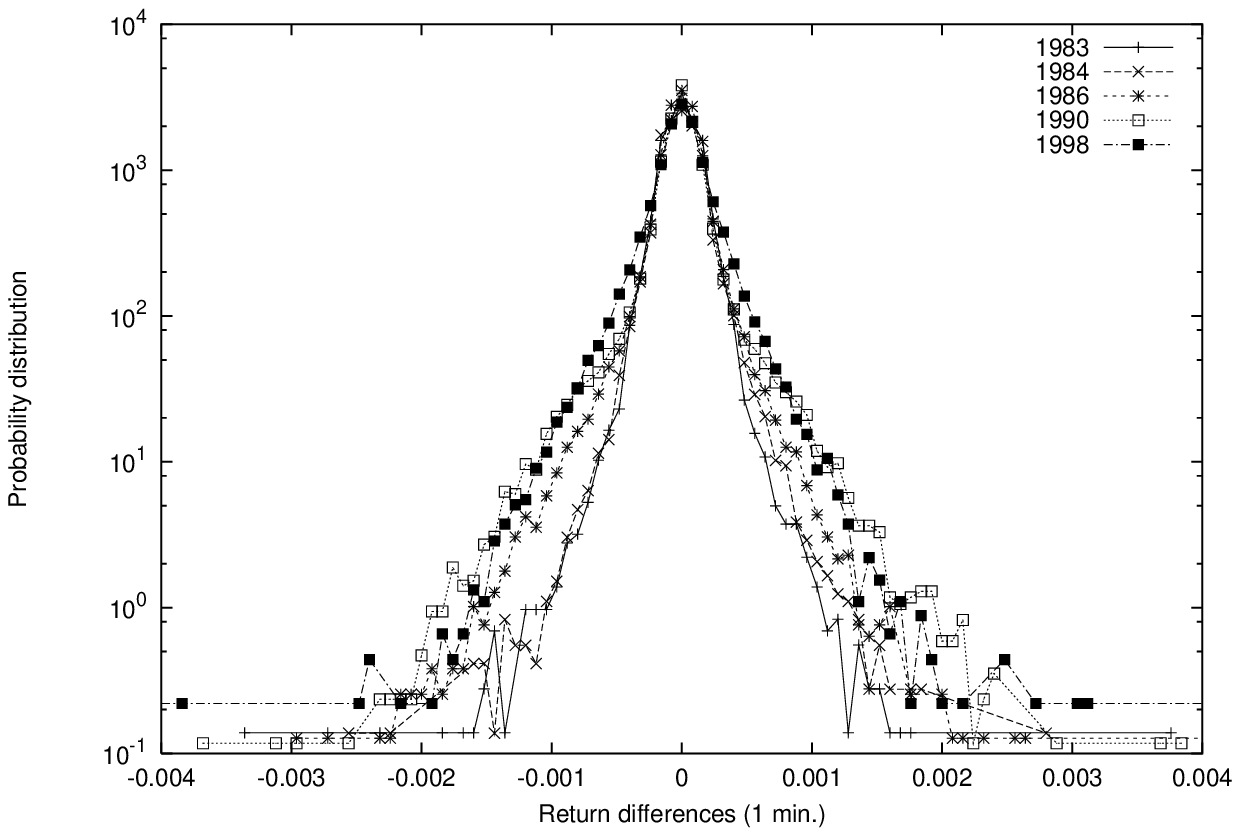}
\end{center}
\caption{We show the empirical pdf's for the return differences of the Standard \& Poor's 500 cash index. First plot involves one-minute returns differences for years
ranging from 1983 to 1999. Second graph is a detail of the previous graph plotting pdf's of years exponentially distributed between 1983 and 1998.}
\label{ret}
\end{figure}

Tick data from the Standard \& Poor's 500 stock cash index is a good example of a multiplicative process, and shows the differences between working in terms of the return $W(t;\tau)$ and in terms of the stock $Z(t;\tau)$. In this case, time series analysis is able to give the empirical probability distributions of the differences between variables which are evaluated at two distinct times.

For instance, stock price differences defined as $Z(t;\tau)=S(t+\tau)-S(t)$ are plotted in Figure~\ref{dif}. The two graphs give the probability distributions in tick data units when $\tau=1$ minute. The first graph shows one-minute stock differences for the seventeen different years, from 1983 to 1999. And the second one only shows stock differences pdfs for the years: 1983, 1984, 1986, 1990, 1998. In this second plot we can see how the wings become fatter as the time increment $t$, in years, increases exponentially.

Data analysis with time series assumes an annual periodicity, in the sense that  $Z(t;\tau) \sim Z(t+\Delta;\tau)$ (in probability) where $\Delta=$ 1 year. In fact, we plot $Z$ ignoring the time $t$ variation and fixing $\tau$ equals to 1 minute. As we have seen from equation~(\ref{phiz}), the probability distribution depends on time $t$, fact which is in contradiction with the assumption that $Z(t;\tau)$ is (in statistical sense) similar to $Z(t+\Delta;\tau)$. We see in figure~\ref{dif} that plots become fatter as time $t$ in years increases.
    
On the other hand, Figure~\ref{ret} shows the return differences defined with the function $W(t;\tau)=R(t+\tau)-R(t)$. We plot the same type of probability distributions of the stock differences but for the return differences. If we compare the two figures we see that, in the returns case, the behavior does not change dramatically over different years ({\it i.e.}, with time $t$) as the case of the stock differences. 
In effect, as it is proved in equation~(\ref{pw}), the return difference, when pdf is that of equation~(\ref{wp}), does not depend on time $t$ and $W(t;\tau)=W(\tau)$.

We could also have plotted the relative stock differences but we do not think to be necessary. Relative difference $Y$ is a better estimating variable than $Z$ but it is not stationary unless we modify their value in an addequate way. We will study this in the next section.

\section{Data analysis and estimators}

This section goes deeper in the study and comparison of $W(t;\tau)$ and $Z(t;\tau)$ estimators. Section~2 and 3 give us all necessary tools for measuring their quality, and thus giving correctly the first and second moment of the stock and return stochastic variables. In this way, we will also study some facts of $Y(t;\tau)$ estimator and see that in some sense this estimator is halfway between the stock and return differences. 

\subsection{Estimators for the first moment of the return and stock}

\begin{figure}
\begin{center}
\epsfbox{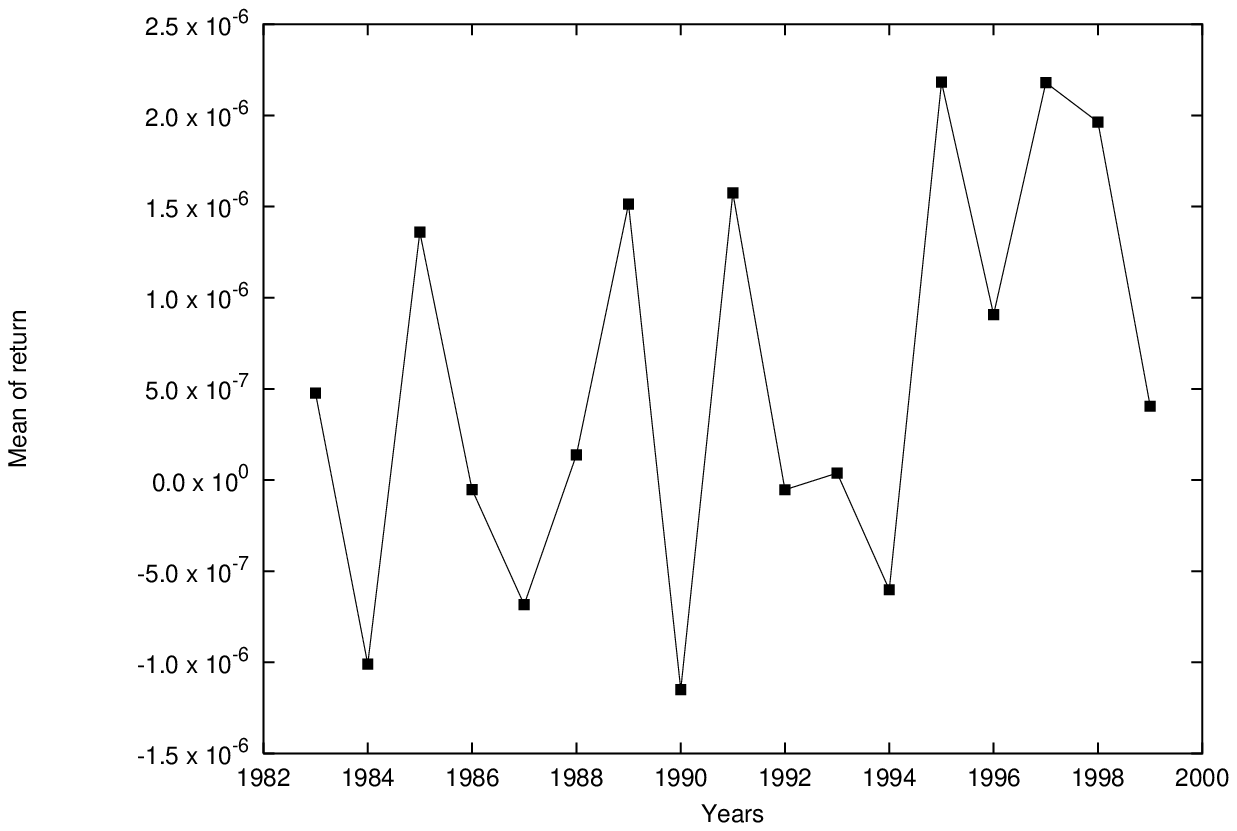}
\epsfbox{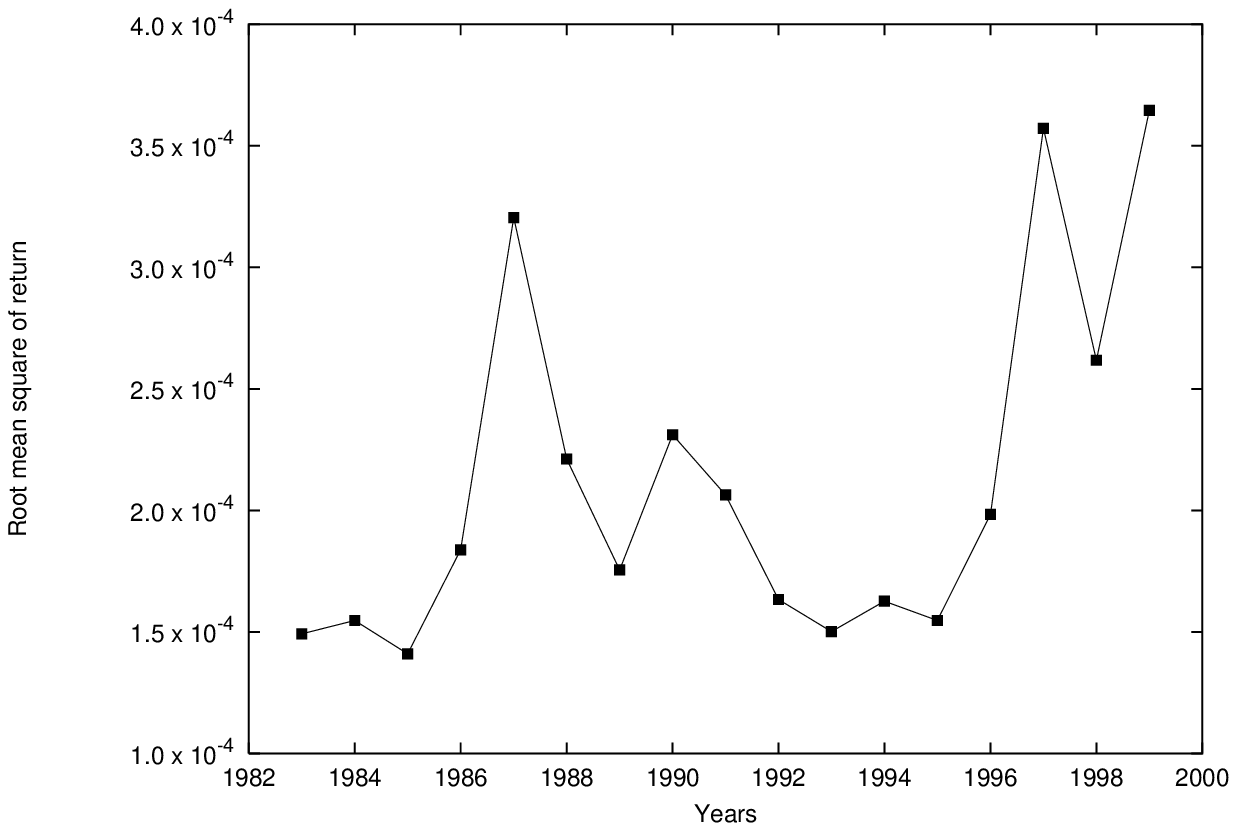}
\end{center}
\caption{First and second moments of the one-minute return differences. We plot respectively $M_W(t;\tau=1,T=1)$ and $[V_W(t;\tau=1,T=1)]^{1/2}$ as a function of time $t$ in years, from 1983 to 1999. Those functions are defined in equations~(\ref{Mw}) and~(\ref{Vw}).}
\label{retmom}
\end{figure}

\begin{figure}
\begin{center}
\epsfbox{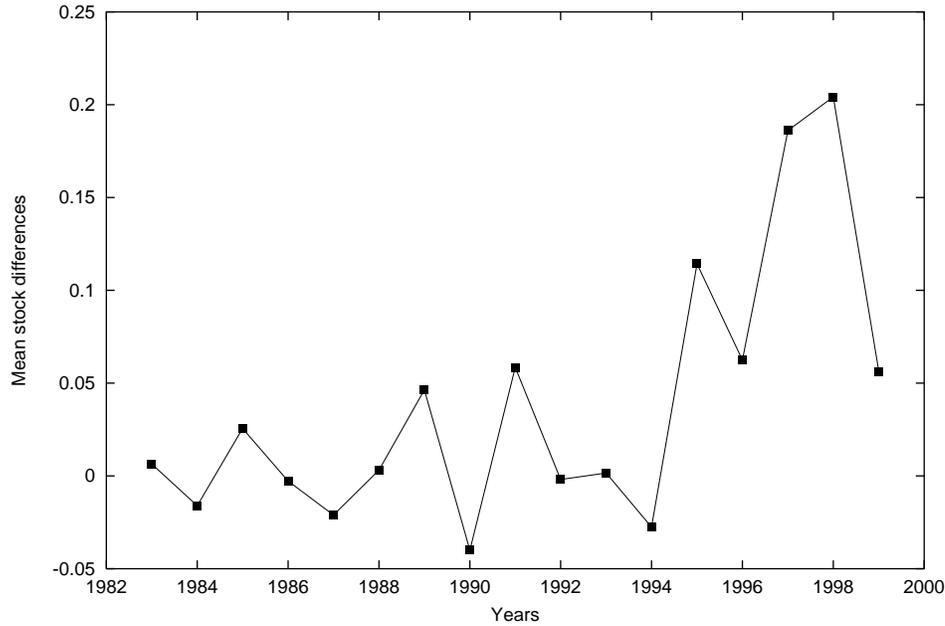}
\end{center}
\caption{First moment estimation with the one-minute stock difference. We plot $Z(t;N\tau=1\mbox{ year})/N$, for estimating $\langle Z(t;\tau=1\mbox{ day})\rangle$, as a function of time $t$  (in years) from 1983 to 1999. Exponential growth with time $t$ is observed as is also shown in equation~(\ref{mt}).}
\label{difmom1}
\end{figure}

We define the following two sums for estimating the first moment:
\begin{eqnarray}
M_W(t;\tau,N) &\equiv& \frac{1}{N} \ \sum_{n=0}^{N-1} W(t+n\tau;\tau)
= \frac{1}{N} \  W(t;N\tau),
\label{Mw}
\\
M_Z(t;\tau,N) &\equiv& \frac{1}{N} \ \sum_{n=0}^{N-1} Z(t+n\tau;\tau)
= \frac{1}{N} \  Z(t;N\tau).
\label{Mz}
\end{eqnarray}
In latter expressions, we have synthesized the sum with the definition of the stock
differences given respectively in equations~(\ref{w}) and~(\ref{Z}). We want to study the quality of those estimators~\cite{cramer}. Their averages are
\begin{equation}
\fl
\langle M_Z(t;\tau,N)\rangle
= \frac{1}{N} \ \langle Z(t;N\tau) \rangle,
\qquad
\mbox{and} 
\qquad
\langle M_W(t;\tau,N) \rangle
= \frac{1}{N} \ \langle W(t;N\tau) \rangle.
\label{ms}
\end{equation}

We can easily derive the average over $M_W$ if we take into account equation~(\ref{pw}). We see there that $W$ evolves in the same way as the return and thus
\[
\langle M_W(t;\tau,N) \rangle= \frac{1}{N} \ \langle R(N\tau) \rangle.
\]
However, this first moment can be decomposed in a sum of $N$ equivalent terms although evaluated at several different times ---see equation~(\ref{<r1-r2>}). That is: 
\begin{equation}
\langle M_W(t;\tau,N) \rangle=  \langle R(\tau) \rangle.
\label{<Mw>}
\end{equation}
Hence, the average of $M_W$ estimator is equal to the first moment for the return. In the first plot of figure~\ref{retmom}, we have $M_W$ estimator for the case when $N\tau=1$ year and $\tau=1$ minute. We observe that first moment changes from one year to another but with any specific trend.

On the other hand, the average over $M_Z$ is obtained according to the expression~(\ref{<z>}), and reads
\[
\langle M_Z(t;\tau,N) \rangle = \frac{1}{N} S_0 \phi_R(-\rmi,t|0) 
[\phi_R(-\rmi,N\tau|0)-1].
\]
And, from equation~(\ref{<s>}), we finally have
\begin{equation}
\langle  M_Z(t;\tau,N) \rangle = 
\frac{\langle S(t) \rangle}{N S_0}
[\langle S(N\tau) \rangle -S_0].
\label{<Mz>s}
\end{equation}
In fact, the estimator is supposed to approach to the following value as the number of sample data increases
\begin{equation}
\lim_{N \rightarrow \infty} \langle  M_Z(t;\tau,N) \rangle 
\longrightarrow  \frac{\langle S(t) \rangle}{S_0} \ 
(\langle S(\tau) \rangle- S_0),
\label{<M>}
\end{equation}
which when $\tau$ is small (keeping only first order contribution) becomes 
\begin{equation}
\lim_{N \rightarrow \infty} \langle  M_Z(t;\tau,N)  \rangle \longrightarrow 
\langle S(t) \rangle \ \langle {R}(\tau) \rangle.
\label{<Mtau>}
\end{equation}
However, note that equations~(\ref{<M>}) and~(\ref{<Mtau>}) will be valid only in case that following and equivalent limits are true
\begin{equation}
\lim_{N\rightarrow \infty} 
\ \frac{\phi_R(-\rmi,N\tau|0)-1}{N \ [\phi(-\rmi,\tau|0)-1]} \longrightarrow 1
\qquad 
\lim_{N\rightarrow \infty} 
\ \frac{\langle S(N\tau) \rangle - S_0}
{N \ [\langle S(\tau) \rangle-S_0]} \longrightarrow 1.
\label{lim}
\end{equation}
Unfortunately this is not true in general. For instance, the Wiener process has the following limits
\begin{equation}
\lim_{N \rightarrow \infty} 
\frac{\rme^{(\mu +\sigma^2/2)N \tau}-1}
{N \left[\rme^{(\mu+\sigma^2/2)\tau}-1\right]} \longrightarrow \infty,
\label{wlim}
\end{equation}
where we take into account equation~(\ref{w<s>}). Although $\tau$ is very small, the term will tend to infinity as $N$ approaches to infinity. Therefore, for this case, {\it the stock differences estimator of the first moment is a biassed and not consistent estimator}~\cite{cramer}. 

We also observe the $\langle M_Z \rangle$ depends on time $t$ and, for the Wiener case, estimator evolves in average as 
\begin{equation}
\langle M_Z(t;\tau,N) \rangle  = S_0 \ \rme^{(\mu+\sigma^2/2)t} 
\ \frac{1}{N} \ \left[ \rme^{(\mu+\sigma^2/2)N\tau}-1 \right], 
\label{mt}
\end{equation}
which is derived taking into account equations~(\ref{w<z>}) and~(\ref{<Mz>s}). We see that the average grows exponentially with time $t$. This phenomena is also empirically observed in figure~\ref{difmom1} for the one-minute stock differences graph. In this case, $\tau=$ 1 minute, $N \tau=T=$ 1 year and $t=kT$ is evaluated in years. Therefore, equation~(\ref{mt}) reads
\[
\langle M_Z(k;\tau= 1 \mbox{min.},N) \rangle  = 
S_0 \ \rme^{(\mu+\sigma^2/2)kT} \
\frac{T}{\tau} \ \left[ \rme^{(\mu+\sigma^2/2)T}-1 \right],
\]
where we can see that average grows exponentially with $k$ similarly to figure~\ref{difmom1} which plots the first moment estimator in terms of $k$ from 1988 to 1999.
 
We may now study the limiting value of $N$ for which $Z$ gives a good estimation of $\langle S(\tau) \rangle$. For this to be possible
\[
\frac{\langle S(N\tau) \rangle - S_0}{N \ [\langle S(\tau)\rangle-S_0]} \sim 1,
\]
which is equivalent to demand
\begin{equation}
\langle Z(t;N\tau) \rangle \sim N \langle Z(t;\tau) \rangle.
\label{N}
\end{equation}
Assuming that the market follows the Wiener process, it can be shown that $M_Z$ gives a good estimation for $\langle S(\tau) \rangle$ when
\begin{equation}
\frac{1}{2}(\mu+\sigma^2/2)N\tau \ll 1.
\label{wN}
\end{equation}

In addition, the variance of the estimator will determine us whether is an efficient estimator or not. Thus, 
\begin{equation}
{\rm Var} [M_Z(t;\tau,N)] 
= 
\frac{1}{N^2} \left[ \langle Z^2(t;N\tau) \rangle
-\langle Z(t,N\tau) \rangle^2 \right].
\end{equation}
Taking into account equations~(\ref{<z>}) and~(\ref{<z^2>}), we finally obtain the variance of the estimator in terms of the characteristic function
\begin{eqnarray}
\frac{1}{S_0^2} {\rm Var} [M_Z(t;\tau,N)]
&=&
\phi_R(-2\rmi,t|0) \left[1+\phi_R(-2\rmi,N\tau|0) -
2 \ \phi_R(-\rmi,N\tau|0)\right] 
\nonumber \\ 
&& \ \ \ \ \ \ -
\phi^2_R(-\rmi,t|0) \ [\phi_R(-\rmi,N\tau|0)-1]^2.
\label{<Mvar>}
\end{eqnarray}
This can be represented in terms of the moments of the stock as
\begin{eqnarray*}
\fl {\rm Var} [M_Z(t;\tau,N)]
&=& \frac{\langle S(t)^2 \rangle}{N^2 \ S_0^2} \left[
S_0^2 +\langle S(N\tau)^2 \rangle  -
2 \ S_0 \ \langle S(N\tau) \rangle \right] 
\\&&- \frac{\langle S(t) \rangle^2}{N^2 \ S_0^2} \  
\left[ \langle S(N\tau) \rangle - S_0 \right]^2.
\end{eqnarray*}
And it is said to be a good ({\it i.e.}, efficient) estimator when its variance tends to zero as $N$ tends to infinity~\cite{cramer}.

For the particular case of the Wiener process we will have
\begin{eqnarray}
\fl {\rm Var} [M_Z(t;\tau,N)]
&=& \frac{S_0^2}{N^2} \ \Bigl\{
\rme^{2(\mu + \sigma^2) t} \left[
1 + \rme^{2(\mu + \sigma^2) N \tau}  -
2 \ \rme^{(\mu + \sigma^2/2) N \tau} \right]
\nonumber \\
&& \ \ \ \ \ \ \ \
- \rme^{2(\mu + \sigma^2/2) t} \
\left[1+\rme^{2(\mu + \sigma^2/2) N \tau}-2 \ 
\rme^{2(\mu + \sigma^2/2) N \tau} \right]
\Bigr\},
\nonumber \\
\label{wMvar}
\end{eqnarray}
which also diverges as $N\rightarrow \infty$. We then conclude that, in general, this estimator is not efficient. For the Wiener case, the estimator is efficient only in case that $N$ is limited by the maximum value given by equation~(\ref{wN}).

\begin{figure}
\begin{center}
\epsfbox{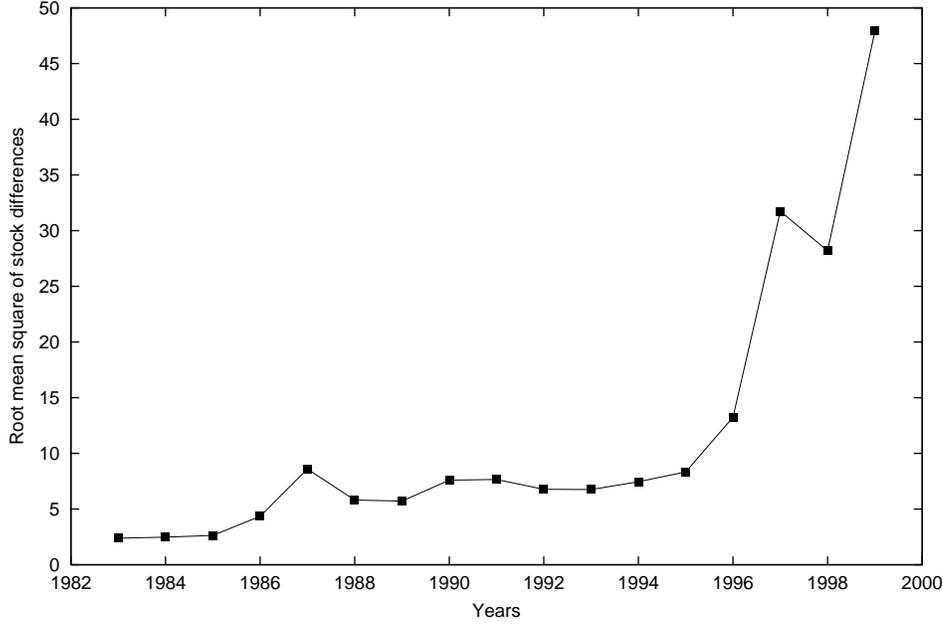}
\end{center}
\caption{Square root for the variance of the one-minute stock differences. We here plot $[V_Z(t;\tau=1,N)]^{1/2}$ as a function of time $t$ in years, ranging from 1983 to 1999. Exponential growth with time $t$ is observed as is also shown in equation~(\ref{m2t}).}
\label{difmom2}
\end{figure}

\subsection{The variances for the stock and the return differences}

We define the estimator for the variance $Z(t;\tau)$ and $W(t;\tau)$ by
\begin{equation}
V_Z(t;\tau,N)=\frac{1}{N-1} \ \sum_{n=0}^{N-1} 
[Z(t+n\tau;\tau)-M_Z(t,\tau;N)]^2.
\label{Vz}
\end{equation}
\begin{equation}
V_W(t;\tau,N)=\frac{1}{N-1} \ \sum_{n=0}^{N-1} 
[W(t+n\tau;\tau)-M_W(t,\tau;N)]^2.
\label{Vw}
\end{equation}
Similarly to the case above, we average the variance estimator in these two cases. Thus we have
\[
\langle V_W(t;\tau,N) \rangle = \frac{1}{N-1} \
\sum_{n=0}^{N-1} \left[
\langle W(t+n\tau,\tau)^2
\rangle \right] - \ \frac{1}{N(N-1)} \langle W(t;N\tau)^2\rangle,
\]
and
\[
\langle V_Z(t;\tau,N) \rangle =
\frac{1}{N-1} \
\sum_{n=0}^{N-1} \left[
\langle Z(t+n\tau,\tau)^2
\rangle \right]
- \frac{1}{N(N-1)} \langle Z(t;N\tau)^2 \rangle,
\]
where we have used equations~(\ref{Mw}) and~(\ref{Mz}). For the return differences case defined in equation~(\ref{w}) and using the properties summarized in equation~(\ref{pw}), we obtain
\begin{equation}
\langle V_W(t;\tau,N) \rangle = \frac{1}{N-1}\left[N \langle R^2(\tau) \rangle -\frac{1}{N} \langle R^2(N\tau)\rangle \right].
\label{<Vw>}
\end{equation} 
But if we particularize to the Wiener case, the variance average is directly related to the variance of the return, {\it i.e.},
\[
\langle V_W(t;\tau,N) \rangle = \sigma^2 \tau,
\]
where we take into account equation~(\ref{w<r>}).

And for the stock differences, we find similar divergences and limiting values for sample data to that of the case of the first moment estimator $M_Z$. Let us show this.
We can write in terms of the characteristic function of the return with the help of equation~(\ref{<z^2>}) the average variance defined above
\begin{eqnarray*}
\fl \langle  V_Z(t;\tau,N) \rangle &=& \frac{S_0^2}{N-1}
\left\{[1+\phi_R(-2\rmi,\tau|0)-2\phi_R(-\rmi,\tau|0)] \sum_{n=0}^{N-1}
\phi_R(-2i,t+n\tau|0) \right.
\nonumber \\
&& \; \; \; \; \; \ \ \
-\frac{1}{N}[1+\phi_R(-2\rmi,N\tau|0)-2 \phi_R(-\rmi,N\tau|0)] \phi_R(-2\rmi,t|0)
\Biggr\}.
\label{<V>}
\end{eqnarray*}
Moreover, in terms of averages over the stock given in equations~(\ref{<s>}) and~(\ref{<s^2>}), we have that the average is 
\begin{eqnarray*}
\fl \langle  V_Z(t;\tau,N) \rangle &=& \frac{1}{N-1} \
\left\{[S_0^2 + \langle S(\tau)^2 \rangle - 2 \ 
S_0 \ \langle S(\tau) \rangle] \sum_{n=0}^{N-1}
\frac{\langle S(t+n\tau)^2 \rangle}{S_0^2}
\right.
\nonumber \\
&&
\ \ \ \ \ \ \ \ \ \ \ \ \ \ \ \ \
\left.
-\frac{1}{N}[S_0^2 + \langle S(N\tau) \rangle^2 - 2 
S_0 \ \langle S(N\tau) \rangle] 
\frac{\langle S(t)^2\rangle}{S_0^2}
\right\}.
\label{<V>b}
\end{eqnarray*}
For the case when market model is the Wiener process, we then have
\begin{eqnarray}
\fl  \langle  V_Z(t;\tau,N) \rangle &=& \frac{S_0^2}{N-1} \rme^{2(\mu+\sigma^2)t}
\left\{ \left[1 + \rme^{2(\mu+\sigma^2)\tau}-2\rme^{(\mu+\sigma^2/2)\tau}\right]
\frac{\rme^{2(\mu+\sigma^2)N\tau}-1}{\rme^{2(\mu+\sigma^2)\tau}-1}
\right.
\nonumber \\
&&
\ \ \ \ \ \ \ \ \ \ \ \ \ \ \ \ \ \ \ \ \ \ \ \ \
-\frac{1}{N}\left[1 + \rme^{2(\mu+\sigma^2)N\tau}-2\rme^{(\mu+\sigma^2/2)N\tau}\right]
\Biggr\}.
\label{w<V>}
\end{eqnarray}
Analogously to the first moment case, as $N$ tends to infinity the average of the estimator diverges. And, the estimator will be valid only when $N$ obeys condition~(\ref{wN}). If this condition holds and keeping $\tau$ small, equation~(\ref{w<V>}) proves that $V_Z$ is a good estimator for the volatility. Hence,
\begin{equation}
\langle V_Z(t;\tau,N) \rangle = S_0^2 \ \rme^{2(\mu+\sigma^2)t} \ \sigma^2 \tau,
\label{m2t}
\end{equation}
where we also observe that this estimator also grows exponentially with time $t$. This phenomena is also empirically observed in figure~\ref{difmom2}. Observe that last expression can be rewritten in terms of the second moment for the stock given by equation~(\ref{w<s^2>}), that is:
\[
\langle  V_Z(t;\tau,N) \rangle= \langle S^2(t) \rangle \ \sigma^2 \tau. 
\]
Hence, for avoiding this divergence with $t$, a possible solution is to consider the estimator in the following way
\[
\frac{\langle  V_Z(t;\tau,N) \rangle}{\langle S^2(t)\rangle }= \sigma^2 \tau.
\]

\section{Conclusions}

Data analysis in financial markets is a very important issue due to the strong demand of higher precision in the estimation of parameters describing markets dynamics. For this reason, we have studied the stock and return differences when the return process is Markovian and homogeneous. Starting from the return characteristic function, we have 
derived the first and second moments, the variance and the correlation for the return, the stock price, and the return and stock differences. We have also obtained these expressions for the particular case when prices are driven by a Wiener process. After these calculations, we have compared the data analysis performed with the return and stock prices differences in the particular case that data source is the Standard \& Poor's 500 cash index.
 
We have intended to stress the importance in the way we manage financial database. It is well-known that stock data follows a multiplicative stochastic process but in some situations was, and still is, preferred to handle stock differences instead of taking return differences, that is: $S(t+\tau)-S(t)$ instead of $R(t+\tau)-R(t)$~\cite{bachelier,bouchaud}. The usual reason for doing this is that when $\tau$ is small one can approximate the logarithm differences with the stock differences. We have showed that in general it is not true since estimators for the stock differences are biassed and not efficient. The approximation will be valid only when the sample data is smaller than a certain ``critical" value and we have obtained a rule for the estimation of this value.

\ack

This work has been supported in part by Direcci\'on General de Proyectos de Investigaci\'on under contract No.BFM2000-0795, and by Generalitat de Catalunya under contract No.2000 SGR-00023. We thank J.M. Porr\`a for showing and stressing us the importance of this backing issue.

\end{document}